\documentclass[proof]{pasj00}     
 \SetRunningHead{K.Ono et al.}{Ono}
\Received{}
\Accepted{}
\Published{}         
\begin{document}
\title{A Suzaku Observation of the Low-Mass X-Ray Binary GS1826-238 in the Hard State} 
\author{Ko ONO$^{1}$, Soki SAKURAI$^{1}$, Zhongli ZHANG$^{1}$, Kazuhiro NAKAZAWA$^{1}$, and Kazuo MAKISHIMA$^{1,2,3}$}
\affil{1 Department of Physics, The University of Tokyo, 7-3-1 Hongo, Bunkyo-ku, Tokyo 113-0033 \par
	   2 Research Center for the Early Universe, The University of Tokyo, 7-3-1 Hongo, Bunkyo-ku, Tokyo 113-0033 \par
	   3 MAXI Team, Global Research Cluster, The Institute of Physical and Chemical Research, 2-1 Hirosawa, Wako, Saitama 351-0198}
\email{ono@juno.phys.s.u-tokyo.ac.jp}
\KeyWords{accretion disks, accretion, stars: neutron, X-rays: binaries}
\maketitle

\draft
\begin{abstract}
The neutron star Low-Mass X-ray Binary GS 1826$-$238 was observed with Suzaku on 2009 October 21, for a total exposure of 103 ksec. Except for the type I bursts, the source intensity was constant to within $\sim 10 \%$. Combining the Suzaku XIS, HXD-PIN and HXD-GSO data, burst-removed persistent emission was detected over the 0.8--100 keV range, at an unabsorbed flux of  $2.6\times 10^{-9}$~erg s$^{-1}$ cm$^{-2}$. Although the implied 0.8--100 keV luminosity, $1.5 \times 10^{37}$ erg s$^{-1}$ (assuming a distance of $7$~kpc), is relatively high, the observed hard spectrum confirms that the source was in the hard state. The spectrum was successfully explained with an emission from a soft standard accretion disk partially Comptonized by a hot electron cloud, and a blackbody emission Comptonized by another hotter electron cloud. These results are compared with those from previous studies, including those on the same source by \citet{Thompson05} and \citet{Cocchi11}, as well as that of Aql X-1 in the hard state obtained with Suzaku \citep{Sakurai14}. 
\end{abstract}

\section{Introduction\label{sec:introduction}}
 A neutron-star (NS) Low Mass X-ray binary (LMXB) is one of the most typical X-ray sources involving NS. It has a low-mass ($\leqq 1M_{\odot}$) companion star, from which it accretes mass. When the mass accretion rate is high, these objects are found in the so-called soft state. The soft-state spectra have long been understood to consist of a multi-color disk blackbody (MCD) emission from a standard accretion disk, and a blackbody radiation from the NS surface \citep{Mitsuda84}. This ``Eastern model'' has been confirmed repeatedly with various observations (e.g., \cite{Makishima89}; \cite{Takahashi08}, \cite{Sakurai14}) to provide a better description of LMXBs, at least in the soft state, than the ``Western model'' \citep{White88} which invokes a blackbody and a Comptonized disk emission.
\par
When the accretion rate falls typically below a few percent of the Eddington limit, these objects are found in the so-called hard state, with spectra apparently harder than  those in the former state. As the hard X-ray sensitivity improved, one of the major objectives of the LMXB study has become to understand their hard state. The spectrum in this state usually shows a power-law like shape with a photon index of $\sim 2$, typically extending up to $\sim 100$~keV. This spectrum is generally interpreted as due to strong Comptonization, because such a radiation spectrum is expected to have a power-law shape extending up to a cutoff energy determined by the electron temperature. In addition, an independent optically-thick emission is often needed to explain spectral excess seen in $\lesssim 2$~keV (\cite{Lin07}; \cite{Tarana11}). However, the origin of the Compton seed photons and the interpretation of the soft-excess component both remained ambiguous.
\par
\citet{Sakurai12} and \citet{Sakurai14} addressed the above questions by analyzing 7 Suzaku data sets of the transient LMXB Aquila X-1, obtained during its outburst in 2007. They successfully explained the broad-band spectra of this source in the hard state considering that the Compton seed photons are provided by the blackbody emission from the NS surface, and that the soft excess is produced by the MCD emission from an accretion disk which is truncated at a radius of $\sim 20$~km, larger than that of the NS. At this radius, the accreting matter is considered to turn into an optically-thin hot flow, i.e. a corona, and plunges onto the NS surface to be thermalized therein. The heated NS surface emits the blackbody photons, which are Comptonized by the subsequent hot flow. They thus succeeded in understanding the accretion geometry of Aql X-1 in the hard state, as a natural extension from the Eastern-model picture developed for the soft state. Our next step is to examine whether or not this understanding generally applies to other LMXBs in the hard state. 
\par
For the above purpose, we chose the LMXB, GS 1826$-$238, which was first discovered by Ginga \citep{Makino88}, and subsequently found to have a low-mass companion star with a magnitude of V=19.3 \citep{Barret95}. This source regularly emits type I X-ray bursts \citep{Barret94,Ubertini97}, which confirm the presence of a neutron star. From the peak fluxes of these type I X-ray bursts, the distance to this source has been constrained as $\leqq 9.6$~kpc. In the present paper, the distance is assumed to be 7 kpc \citep{Barret00}. It has so far been found usually in the hard state, and is hence suited for our purpose. As a particularly interesting aspect of this source, \citet{Thompson05} and \citet{Cocchi11} already studied its Chandra, RXTE, and BeppoSAX spectra, and constructed a view that not only the blackbody but also the disk emission is strongly Comptonized; this view in some sense resembles the Western model. In the present study, we keep these works also in mind.\par

\section{Observation and Data Reduction}

\subsection{Observation}
 We used an archival Suzaku data set of GS1826$-$238 (ObsID 404007010). The observation was performed on 2009 October 21 from 20:22:19 UT for a gross duration of 184 ks and a total exposure of $\sim 103$~ks, using the XIS and the HXD onboard. The source was placed at the ``HXD nominal'' position. In order to avoid event pile up, the XIS was operated in ``1/4 window mode'', wherein the CCD events are read out every $2$~s. 

\subsection{XIS data reduction} \label{xis_data_reduction}
 The present paper utilizes XIS0 events of GRADE 0, 2, 3, 4, and 6. We accumulated on-source XIS events over an entire region of the XIS0 image. Figure 1a shows 0.5--10~keV light curve from XIS0. Out of several bursts which occured in the observation period, the XIS light curve reveals 6 events that survived our data screening criteria. After excluding all these bursts, the XIS0 count rate was consistent, within $\sim10\%$, with a constant 17.8 cts s$^{-1}$. Therefore, we created an XIS on-source spectrum by accumulating all the XIS events from a circular region of radius $2.4'$, but excluding the Type I bursts (typically $\sim 400$~s each). Furthermore, to avoid pile-up effects, we eliminated the image center within a radius of $1'$. The background events were obtained over a circular region which does not overlap with the annular source region and were subtracted. By discarding the image center, the 0.5--10~keV signal rate decreased to $7.57\pm0.01$ counts s$^{-1}$. The obtained XIS0 spectrum is shown in figure \ref{fitting_a} (black).
 
\begin{figure}[h]
	\begin{center}
		\FigureFile(80mm,50mm){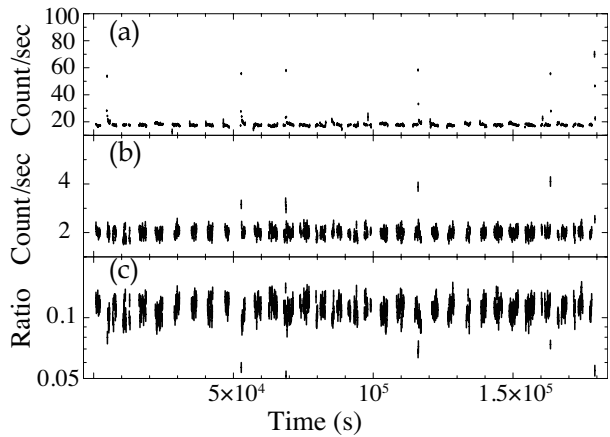}
   \end{center}
   \caption{Light curves from XIS0 (panel a; $0.5-10$~keV) and HXD-PIN (panel b; 16--60 keV), with 128 sec binning. Background was subtracted from the latter, but not from the former since it is negligible ($< 8\%$). Panel (c) shows the hardness ratio of the XIS0 to HXD-PIN count rates. Time 0 corresponds to 2009 October 21, UT 20:22:19.}\label{fig:light_curve}
\end{figure}

\subsection{HXD data reduction} \label{sec:hxd_data_reduction}
In the same way as the XIS0 data, we accumulated the cleaned HXD-PIN events over the entire exposure, but excluding the Type I bursts, to achieve a net exposure of 78.7 ks (dead time corrected). The simulated Non X-ray Background (NXB) events provided by the HXD team were used to construct an NXB spectrum \citep{Fukazawa09}, which was then subtracted. The Cosmic X-ray Background (CXB) was effectively excluded from the data by adding a fixed CXB model when we fit the spectra. Figure \ref{fig:light_curve}b and c show the background-subtracted HXD-PIN light curve and its ratio to the XIS0 count rates. After subtracting the NXB and performing dead-time correction, the count rates from HXD-PIN (16--60~keV) and HXD-GSO (60--100~keV) were $1.87\pm0.005$~counts s$^{-1}$ and $0.42 \pm 0.01$~counts s$^{-1}$, respectively, where the errors refer to statistical $1$~$\sigma$ uncertainties. Except the bursts, the HXD-PIN count rate was approximately constant. Below, we use HXD data up to 100~keV, where the signal intensity, $\sim4 \times 10^{-5}$ cts cm$^{-2}$ s$^{-1}$ keV$^{-1}$, still exceeds $1\sigma$ systematic error ($\lesssim 7\times10^{-6}$ cts cm$^{-2}$ s$^{-1}$ keV$^{-1}$ for $> 10$~ks; \cite{Fukazawa09}). The derived HXD-PIN and HXD-GSO spectra are shown in figure \ref{fitting_a}, together with that from XIS0.

\section{Spectral Analysis}
As seen in figure \ref{fitting_a}, the source exhibited a hard power-law like spectrum extending up to $100$~keV. Since this is a typical feature of an LMXB in the hard state, we regard the source as in the hard state. We followed the analysis process of \citet{Sakurai12} on Aql X-1.
In order to account for the CXB contribution in the HXD-PIN data (section \ref{sec:hxd_data_reduction}), we expressed it as an analytical model of \citet{Boldt87} as\par
${\rm CXB}(E)=9.41 \times 10^{-3}(\frac{E}{1 \rm keV})^{-1.29}{\rm exp}(-\frac{E}{40 \rm keV})$,
\\ where the unit is ${\rm photons\ cm^{-2}\ s^{-1} keV^{-1} FOV^{-1}}$, and $E$ is the energy in keV. Then, over an energy range of 0.8--100~keV, we fitted simultaneously the XIS0, HXD-PIN, and HXD-GSO spectra as prepared in section 2. Two energy ranges, 1.7--1.9~keV and 2.2--2.4~keV, were excluded from the XIS data, to avoid calibration uncertainties associated with the silicon K-edge and gold M-edge, respectively. Since the XIS image center was excluded (section \ref{xis_data_reduction}), cross normalization between XIS0 and HXD-PIN was adjusted by applying a constant multiplicative factor to the model, and leaving it free for the XIS while fixing it at 1.18 for the HXD \citep{Kokubun07}.

\par
\subsection{A single Comptonized blackbody (Model 0)}
 We selected an XSPEC model {\tt nthcomp} \citep{Zdziarski96,Zycki99} to express the Comptonized component, since it allows us to choose a seed photon source between blackbody and disk blackbody. Just to reconfirm the analysis steps by \citet{Sakurai14}, the spectra were first fitted with a model consisting only of a Comptonized blackbody (hereafter BB) component. We name it Model 0 and  express it as {\tt nthcomp [BB]}, where BB in the bracket means that the seed photons are provided by blackbody. Free parameters are the absorption column density $N_{\rm H}$, the blackbody temperature $T_{\rm BB}$ from the NS, the coronal electron temperature $T_{\rm e}$, optical depth $\tau$ of the corona, normalization of the {\tt nthcomp} component, and the XIS cross normalization constant mentioned above.
 \par
 As shown in figure \ref{fitting_a}a, the spectra were approximately reproduced with $\chi^2_{\nu}(\nu)=1.5\, (297)$; the data and the model agree well below $40$~keV. However, significant positive residuals are seen in $>40$~keV, and in $\lesssim2$~keV to a lesser extent. Thus, a single Comptonized blackbody alone is not enough to reproduce the data over the broad energy band. The hard band residuals suggest inadequate modeling of the Comptonization component. Following \citet{Sakurai12}, we begin with adding another soft optically-thick component to the model, so that the Comptonization model can have additional freedom.

\subsection{Single-source Comptonization plus a soft thermal component (Model 1)}
We hence added a disk blackbody component to Model 0, to construct  Model 1 = {\tt diskbb + nthcomp [BB]} as used in \citet{Sakurai12} and \citet{Sakurai14}. In addition to the 6 free parameters in Model 0, the inner disk temperature $T_{\rm in}$ and the normalization of {\tt diskbb} were left free. To explain possible Fe-K emission line from the disk, which is often broadened \citep{Cackett09}, we further incorporated a Gaussian, and left free its width and the normalization. The Gaussian center energy was first left free, but it became $\sim 6.1$~keV which is unphyscical. Therefore, we fixed it at $6.4$~keV or $6.7$~keV, corresponding to (nearly) neutral or He-like iron atoms, respectively. Then, the case of 6.7 keV gave a worse fit by $\Delta \chi^2=4.9$ than that of 6.4 keV. Therefore, we fix the center energy hereafter at 6.4 keV. As shown in figure \ref{fitting_a}b, this model has improved the fit over that with Model 1, giving $\chi^2_{\nu}(\nu)=1.14\,(293)$. The obtained best-fit model parameters are listed in table \ref{table:parameters}, where the inner disk radius $R_{\rm in}$ was modified from the raw value implied by the {\tt diskbb} normalization by multiplying with a factor $\xi \kappa^{2} = 1.19$ \citep{Kubota98,Makishima00}, where $\xi = 0.412$ is a correction factor for inner boundary condition of the disk, and $\kappa = 1.7$ is a color hardening factor of {\tt diskbb}. The disk inclination was assumed to be $\theta = 62^{\circ}.5$ \citep{Mescheryakov11}. The spectrum was reproduced better; the residuals below $\sim10$~keV were explained away by the addition of {\tt diskbb}, and the fit in $>20$~keV was much improved by an increase in $kT_{\rm e}$. However, noticeable residuals still remain around 20--60 keV.
\par
In an attempt to eliminate the residual structure at 20--60~keV, reflection of the Comptonization component by the disk surface was added to Model 1, using a convolution model, {\tt reflect} (Magdziarz $\&$ Zdziarski 1995). Abundances of all the elements in the disk were fixed to the solar values. However, the fit did not improve significantly, giving $\chi^2_{\nu}(\nu)=1.13 (292)$; the residuals were still left at 20--60~keV.
\par
Following Sakurai et al (2012), we also tested another model by exchanging the seed photon source and the directly seen thermal emission, between the disk and the blackbody. The model is expressed as Model 2 = {\tt BB + nthcomp [diskbb]} with the same free parameters as in Model 1. As listed on the right row in table \ref{table:parameters}, the fit goodness turned out to be nearly the same as that with Model 1, $\chi^2_{\nu}(\nu)=1.10\,(293)$, but the result is less physical, since $R_{\rm in}$ ($4.2$~km) is too small. Therefore, Model 2 is no longer considered hereafter.

\subsection{Two Comptonized blackbodies and a disk blackbody (Model 3)}\label{model3_analysis}
Inspection of figures \ref{fitting_a}b, \ref{fitting_a}c and \ref{fitting_a}d suggest that the hard X-ray residuals from Models 1 and 2 arise because the data turn off more gradually than is predicted by a single-$T_{\rm e}$ Comptonization. In other words, there may be more than one Comptonization component with different electron temperatures, since a corona may not be necessarily isothermal. Actually, \citet{Thompson05} applied such a ``double Comptonization"  modeling to the spectra of GS 1826-238. Although they considered that two coronae have different seed photon sources, we tentatively assume here that a fraction of the blackbody from the NS surface is Comptonized by a hotter corona, while the rest by a cooler one. This leads to Model 3 = {\tt nthcomp (BB) + nthcomp (BB) + diskbb}, where the two {\tt nthcomp} components are allowed to have different $T_{\rm e}$ and different $\tau$ but are constrained to have the same seed $T_{\rm BB}$.
\par
As shown in figure \ref{fitting_a}e, the high-energy spectral shape was successfully explained by this Model 3, and the fit became acceptable with $\chi^2_{\nu}(\nu)=1.01\,(290)$. The fit required an optically-thick ($\tau=16$) and cool ($T_{\rm e}=6.8$~keV) corona, together with an optically-thin ($\tau=1.1$) and relatively hot ($T_{\rm e} > 59$~keV) corona. However, we are still left with a serious problem: the derived value of $R_{\rm in}=5.9\pm^{4.8}_{2.2}$~km (table \ref{table:parameters2} left row) is too small compared to $R_{\rm BB}$ ($7.8\pm^{2.9}_{3.0}$~km) and the typical NS radius. This problem, which already existed in the Model 1 fit, is presumably due to too high a temperature of the disk. It hence suggests that the inner part of the disk is also Comptonized weakly, to acquire a significantly higher color temperature.

\subsection{A Comptonized blackbody and a partially Comptonized disk blackbody (Model 4)}
In section \ref{model3_analysis}, we found on one hand that the data suggest the presence of two Comptonizing coronae, or double Comptonization configuration. On the other hand, the disk emission may also be Comptonized at least partially. Then, the simplest scenario to satisfy these two requirements would be to identify the second (cooler) corona with that scattering the disk photons, rather than a fraction of the BB photons, because this double Comptonization property suggested by the broadband data are relatively insensitive to the seed photon temperature. To describe this condition, we assumed that the NS emission is Comptonized by a single corona (like in Model 1 but unlike Model 3), while expressed the disk Comptonization by an XSPEC model {\tt dkbbfth} \citep{Done06,Hori14}, which assumes that the disk emission is Comptonized (using the {\tt nthcomp} code) from $R_{\rm in}$ up to a larger radius $R_{\rm out}$. The disk outside $R_{\rm out}$ is assumed to be directly visible.
\par
\begin{figure}[h]
	\begin{center}
	         \FigureFile(160mm,50mm){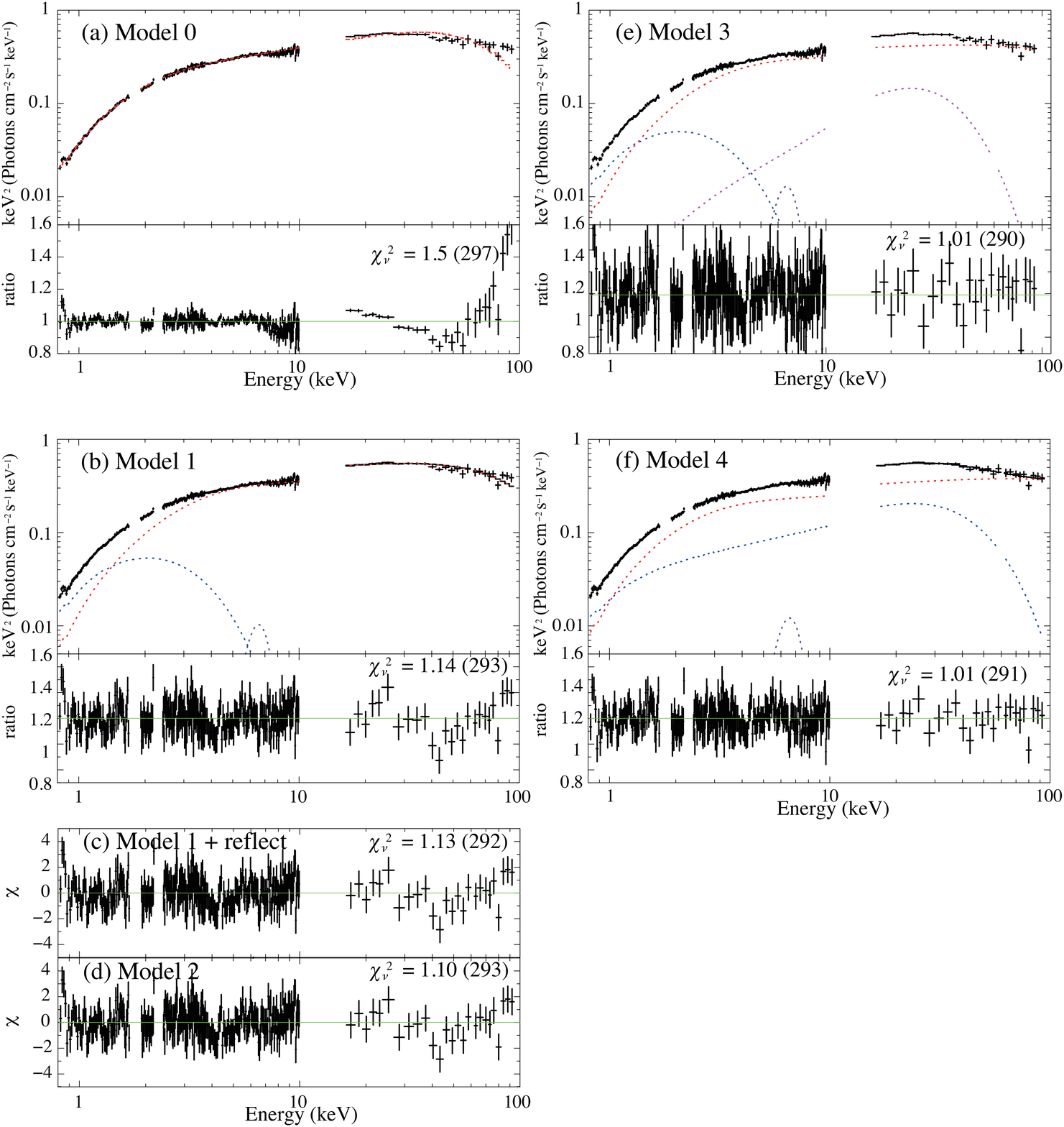}
	   \end{center}
	   \caption{Simultaneous fitting of the XIS0, HXD-PIN and HXD-GSO spectra in $\nu F \nu$ form. (a) A fit with a single Comptonized blackbody (Model 0), and the associated data-to-model ratio. (b) A fit with Model 1, that is, Model 0 plus diskbb. (c) The same as (b) but a reflection is added. (d) Residual of the fit with the same model as (b), but assuming the seed photons are supplied by the disk while the blackbody produces the soft excess. This is Model 2. (e) A fit with two {\tt nthcomp} components and {\tt diskbb} (Model 3). (f) A fit with {\tt nthcomp} assuming a blackbody seed-photon source, and  {\tt dkbbfth}. Red dashed lines in (a), (b), (e) and (f) show the {\tt nthcomp (BB)} components.}\label{fitting_a}
\end{figure}

Utilizing dkbbfth, we constructed Model 4 = {\tt nthcomp (BB) + dkbbfth}. This formalism is in between that of \citet{Sakurai12} and \citet{Sakurai14} which corresponds to $R_{\rm in}\rightarrow R_{\rm out}$, and that of \citet{Thompson05} and \citet{Cocchi11} which is equivalent to $R_{\rm out}\rightarrow \infty$. Since absorption became rather unconstrained due to strong coupling with {\tt dkbbfth}, the column density was fixed to $N_{\rm H}=0.28 \times 10^{22}$~cm$^{-2}$, as obtained with Model 1. The free parameters are the same as in Model 3, plus $R_{\rm out}$. As shown in figure \ref{fitting_a}f and table \ref{table:parameters2}, this model is as successful as Model 3, and gave $\chi^2_{\nu}(\nu)=1.01\,(290)$. Furthermore, as expected, the derived model parameters have become physically reasonable, including $R_{\rm in}>21$~km and $R_{\rm BB}=11.9$~km. While the coronal temperature affecting the disk emission was obtained as $\sim9$~keV (with large errors), that for the NS emission was constrained only as $>50$~keV. This lower limit is still consistent with the value for Aql X-1, $48\pm 6$~keV (Sakurai et al 2014). Thus, we regard Model 4 as our best solution.
\par

 \begin{table}[htb]
 \begin{center}
 \begin{minipage}{10cm}
 \caption{Fit parameters with Model 1 and Model 2.}
  \begin{tabular}{lccc}
   \hline
   \multicolumn{1}{c}{Component}  & parameter							& Model 1					& Model 2	\\
   \hline
   constant			&											& $0.87$					& $0.86$ \\
	   $N_{\rm H}$~$\rm (10^{22}cm^{-2})$&							& $0.28$					& $0.27$  \\
   Opt.thick		& $ T_{\rm BB}/T_{\rm in}$~(keV)        				&$0.75 \pm ^{0.19}_{0.15}$	& $0.70 \pm ^{0.02}_{0.03}$ \\
     	               		& $ R_{\rm BB}/R_{\rm in}$~(km))\footnotemark[$*$ ]	& $7.0 \pm ^{1.9}_{1.4}$		& $4.2\pm^{0.5}_{0.6}$ \\
   {\tt gaussian}		& Sigma~(keV)									&$0.68\pm^{0.34}_{0.25}$		& $0.71 \pm^{0.34}_{0.24}$ \\
   {\tt nthcomp} 		& seed        									& BB						&  disk \\
	                     	& $ T_{\rm in}/T_{\rm BB}$~(keV)					& $0.86 \pm ^{0.22}_{0.12}$	& $1.6 \pm ^{0.15}_{0.13}$\\
    	                 	& $ R_{\rm in}/R_{\rm BB}$~(km))\footnotemark[$*$ ]	& $7.9\pm^{1.3}_{1.7}$ 		& $3.0$\\
           	          	& $ T_{\rm e}$~(keV)							& $26 \pm ^{4}_{3} $			& $31\pm ^{7}_{4}$\\
                     		& $\tau$										& 3.0						& 2.6\\
   \hline
   fit goodness & $\rm \chi ^{2}_{\nu}(\nu)$ & 1.14 (293) & 1.10 (293)\\
   \hline 
\multicolumn{4}{@{}l@{}}{\hbox to 0pt{\parbox{180mm}{\footnotesize
\par\noindent
\footnotemark[$*$] Calculated assuming the source distance of 7 kpc and the inclination angle of 62.5$^{\circ}$.
}\hss}}
   \end{tabular}
   \label{table:parameters}

   \end{minipage}
 \end{center}
 \end{table}

  \begin{table}[htbp]
 \begin{center}
 \caption{Fit parameters with Model 3 and Model 4.}
 \begin{minipage}{10cm}
  \begin{tabular}{lccc}
   \hline
   \multicolumn{1}{c}{Component}  & parameter									& Model 3					& Model 4				\\
   \hline
   constant			&													& $0.99$					& $0.96$		\\
   $\ N_{\rm H}$~$\rm (10^{22} cm^{-2})$  	&									& $0.27$					& 	0.28					\\
   {\tt diskbb}				&$ T_{\rm in}$~(keV)							& $0.78\pm^{0.47}_{0.49}$	&	-					\\
                     				& $ R_{\rm in}$~(km)\footnotemark[$*$ ]				& $5.9\pm ^{4.8}_{2.2}$		& 	-					\\
  {\tt nthcomp/dkbbfth}		&$T_{\rm BB}$/$ T_{\rm in}$~(keV)					& $0.82$					& $0.42 \pm ^{0.08}_{0.20}$	\\
                     				& $R_{\rm BB}$/$ R_{\rm in}$~(km)\footnotemark[$*$ ]	& $2.5\pm^{5.5}_{2.0}$		& $>21$					\\
                     				& $ R_{\rm out}~$(km)							&-						& $>50$					\\
                     				& $ T_{\rm e}$~(keV)							& $6.8\pm^{1.2}_{1.3}$		& $8.9\pm 6$				\\
                     				& $\tau$										& $16$					& 7.6						\\
   {\tt gaussian}				& Sigma~(keV)									& $0.75\pm^{0.31}_{0.22}$	&	$0.73\pm^{0.30}_{0.21}$	\\
   {\tt nthcomp (BB)}			& $ T_{\rm BB}$~(keV)							& -						& $0.63 \pm ^{0.01}_{0.02}$	\\
                     				& $ R_{\rm BB}$~(km)\footnotemark[$*$ ]				& $7.8\pm^{2.9}_{3.0}$		& $11.9\pm 0.3$			\\
                     				& $ T_{\rm e}$~(keV)							& 105 ($>$59)				& $>50$					\\
                     				& $\tau$										& 1.1						& $<1.9$					\\
   \hline
   fit goodness &$\rm \chi ^{2}_{\nu}(\nu)$											& 1.01 (290)				& 1.01 (291)				 \\
   \hline
\multicolumn{4}{@{}l@{}}{\hbox to 0pt{\parbox{180mm}{\footnotesize
\par\noindent
\footnotemark[$*$] Calculated assuming the source distance of 7 kpc and the inclination angle of 62.5$^{\circ}$.
}\hss}}
   \end{tabular}
   \label{table:parameters2}
   \end{minipage}
 \end{center}
 \end{table}

\clearpage

\section{Discussion\label{sec:Discussion}}

We analyzed the Suzaku data of GS1826$-$238 acquired on 2009 October 21. The light curve obtained with XIS0 did not vary significantly, except type I bursts. Using all the data but excluding these bursts, we obtained the 0.8--100~keV spectrum which shows a typical shape of the hard state of LMXBs. After testing several spectral models with progressive complexity, Model 4, {\tt nthcomp (BB) + dkbbfth}, has been found to provide the best representation of the high-quality Suzaku spectrum. The 0.8--100~keV unabsorbed luminosity derived from Model 4 is $L=1.5\times 10^{37}$~erg s$^{-1}$, or $\sim$10\% of the Eddington limit. It places this source at close to the highest-luminosity end of the hard state of LMXBs \citep{Egron13}.
\par
\begin{figure}[htb]
   \begin{center}
         \FigureFile(120mm,35mm){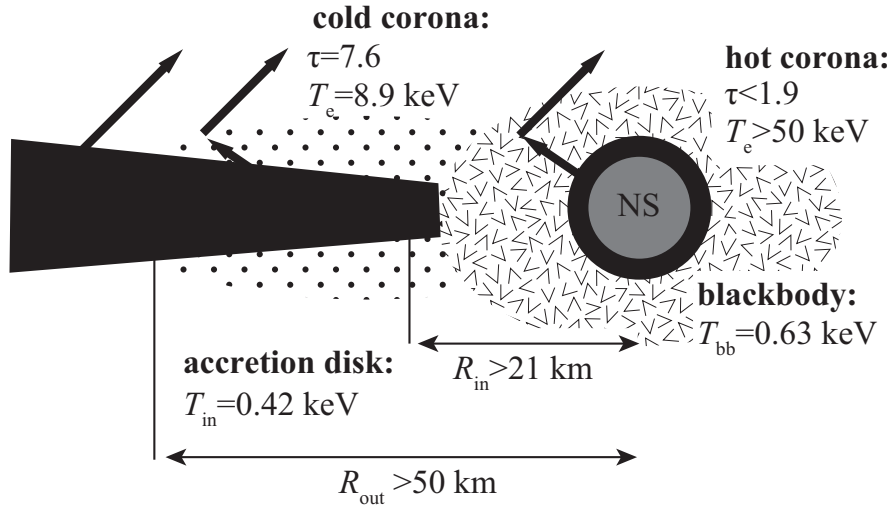}
   \end{center}
   \caption{A schematic cross-sectional view of the accretion flow suggested by Model 4. The two optically-thick emission regions, the disk and the NS surface, are indicated by black, whereas  the two coronae are shown in patterns.}\label{geometry}
\end{figure}

Model 4 consists of two thermal components, a disk blackbody and a blackbody, Comptonized by different coronae. Of the two coronae, the one which Comptonizes the disk emission, assumed to cover an inner part ($<R_{\rm out}$) of the disk, has been found to have a rather low temperature as $8.9$~keV and a high optical depth as 7.6. It is hence suggested to have a rather low scale height above and below the disk, and is probably under strong Compton cooling compared with the hotter corona. Thus, a geometry as illustrated in figure \ref{geometry} can be considered. After passing through the disk-corona coexisting region from $R_{\rm out}$ to $R_{\rm in}$, the accretion flow as a whole becomes a hotter ($T_{\rm e}>50$~keV) and optically-thin coronal stream falling almost spherically onto the NS surface. The matter will then become thermalized on the NS surface, emitting blackbody photons from whole the NS surface ($R_{\rm BB} \sim R_{\rm NS}$). These photons are then Comptonized by the subsequent hot corona, to form a major fraction of the hard X-ray continuum.

\clearpage

\par
In order to assess the physical consistency of Model 4, we calculated the 0.8-100~keV luminosities of the involved model components, and show the results in table \ref{table:luminosities}. There, the disk-related values were corrected for the inclination, while the others are not. Thus, the {\tt dkbbfth} component, namely, the sum of the ``Disk" and ``cooler corona" contributions in table \ref{table:luminosities}, is inferred to carry a 0.8--100 luminosity of $L_1 =0.59 \times 10^{37}$ erg s$^{-1}$. At the same time, the disk is likely to be truncated at $>21$ km, or $>(1.5-1.75) R_{\rm NS}$. Then, ignoring the internal and radial-kinetic energies of the cooler corona, virial theorem predicts that  the NS-related luminosity $L_2$, i.e., the blackbody and ``hotter corona'' contributions summed together, should  be at least $L_2 = 2.0 L_1 = 1.2 \times 10^{37}$ erg s$^{-1}$, which consists of $\sim L_1$ accounting for the remaining half of the energy release from infinity to $ \sim 1.5R_{\rm NS}$, and the full energy output from $\sim 1.5 R_{\rm NS}$ to $R_{\rm NS}$. When this  $L_2$ is fully thermalized on the entire NS surface, we expect to observe a BB temperature of $T_{\rm BB} \gtrsim 0.87$ keV. Compared to this prediction, the actually observed value of $T_{\rm BB}= 0.63$ keV is significantly lower, primarily because $L_2$  is shared between the BB and the hotter corona components. However,  apart from this detail, the measured value of $L_2 = 1.1 \times 10^{37}$ erg s$^{-1}$ (table 3) is lower than the above prediction by at least 10\%, or by $> 0.12 \times 10^{37}$ erg s$^{-1}$. This deficit would increase if considering the neglected energy flows carried by the cooler corona.
\par
One of possible causes of the above discrepancy could be the neglected luminosity of the Comptonized BB above 100 keV, 
which can amount to $0.34 L_2$ and would be sufficient. Another cause could be partial obscuration of the BB component by the disk; an obscured fraction of $\sim 10\%$ would be sufficient.
Yet another possibility is that the missing luminosity is consumed in producing outflows or jets, or in spinning up the NS. If the latter scenario is adopted, the NS in GS 1826$-$238 would spin up from 0 Hz to $\sim$360 Hz in about $10^8 $ yr. Finally, the {\tt dkbbfth} model could be still inaccurate, so that the actual disk radius could be smaller; a value of $R_{\rm in}=16-20$ km would be sufficient to explain away the discrepancy.

\par
Let us compare our results with those from the previous studies. The difference from the model used in \citet{Sakurai12} is the disk Comptonization at $<R_{\rm out}$. Admittedly, the present data gave only a lower limit as $R_{\rm out}>50$~km. This allows a case of the whole disk Comptonization ($R_{\rm out} \rightarrow \infty$), thus making the present result appear closer to those employed by \citet{Thompson05} and \citet{Cocchi11}. Actually, the spectrum analyzed here can be fitted similarly well, with $\chi^2=1.01 (292)$, when replacing {\tt dkbbfth} in Model 4 with {\tt nthcomp[diskbb]} which implies Comptonization of the entire disk. However, by doing so, the inner disk radius increases to $R_{\rm in}=55$~km, and the inner disk temperature decreased to $T_{\rm in}=0.20$~keV, without significant changes in the flux of the Comptonized disk emission. Since this would enlarge the discrepancy between the measured $R_{\rm in}$ and that predicted by virial theorem, the use of the {\tt dkbbfth} modeling is considered to be physically more reasonable. Furthermore, the modeling by \citet{Thompson05} and \citet{Cocchi11} is distinct from ours (regardless of the {\tt dkbbfth/nthcomp[diskbb]} ambiguity), because their fits imply that the Comptonized disk emission accounts for $>50\%$ of the total luminosity, in a larger deviation from the virial theorem requirement. In short, the Comptonizing corona is likely to cover only a limited inner region of the X-ray emitting region of the accretion disk.

\begin{table*}[htb]
	\begin{center}
		\caption{Luminosities of individual components in Model 4.} 
		\begin{tabular}{lcc}

\hline
\shortstack{\\component}							&\shortstack{Luminosity\\ (0.8--100~keV, $\times10^{36}$}~erg s$^{-1}$)	&\shortstack{Fraction\\ (\%)}	\tabularnewline \hline \hline
Disk	\footnotemark[$* \dagger$ ]					&$3.22\pm ^{0.05}_{0.62}$								&$19.4\pm^{0.3}_{3.8}$		\\
Cooler corona\footnotemark[$\dagger \ddagger$]		&$2.72\pm ^{0.27}_{0.35}$								&$16.3\pm^{1.7}_{2.1}$		\\  \hline
Blackbody\footnotemark[$\S$]						&$2.7\pm^{0.29}_{0.02}$									&$16.2\pm^{1.8}_{0.1}$		\\
Hotter corona\footnotemark[$\l$]					&$8.0\pm^{0.71}_{0.18}$									&$48.1\pm^{4.2}_{1.1}$		\\ \hline

\multicolumn{3}{@{}l@{}}{\hbox to 0pt{\parbox{180mm}{\footnotesize
\par\noindent
\footnotemark[$*$] Corrected for the inclination by a factor of $1/\cos62.5^{\circ}$
\par\noindent
\footnotemark[$\dagger$] Sum of the directly visible emission and the seed-photon contribution.
\par\noindent
\footnotemark[$\ddagger$] The luminosity added to the disk photons by the cooler corona.
\par\noindent
\footnotemark[$\S$] The seed blackbody luminosity from the NS surface.
\par\noindent
\footnotemark[$\l$] The luminosity added to the NS-surface emission by the hotter corona.

}\hss}}
		\end{tabular}
		\label{table:luminosities}
	\end{center}
\end{table*}

\par
From the above discussion, we regard the present result as a natural extension of the view by \citet{Sakurai14}, but we need to consider the origin of the clear difference; absence and presence of the inner-disk Comptonization. One possible origin of this difference is in the luminosity. In fact, one of the characteristics of GS 1826$-$238 is its high luminosity ($\sim 0.1 L_{\rm Edd}$) for the hard state of LMXBs. In this regard, the two more LMXBs, 4U 1915$-$05 and MAXI J0556$-$332, are reported to be in the double Comptonization condition in the soft state \citep{Zhang14,Sugizaki13}. Therefore, It is possible that the optically-thick corona on the disk surface starts growing, e.g., as if evaporating from the disk, when an LMXB in the hard state becomes very luminous, comparable to the soft-state luminosities. Indeed, Aql X-1 also required the double Comptonization modeling (Sakurai 2015) when it was at the highest-luminosity end ($1.5\times 10^{37}$ erg s$^{-1}$) of the hard state just before it made a transition into the soft state. Such a cool corona considered here, which covers the disk with a larger optical depth and presumably with a low scale height, could be identified by those proposed by \citet{Kawaguchi01}. In addition, there can be other possibilities, including some effects of inclination, or weak magnetic fields.

\section{Conclusion}
We analyzed an archival data set of GS 1826$-$238 taken on 2009 October 21. Although the luminosity at 0.8--100~keV was rather high at $1.5\times10^{37}$erg s$^{-1}$, the source was in the hard state. The 0.8--100~keV persistent spectrum of GS 1826$-$238 was explained successfully by a disk blackbody partially Comptonized by a cool and optically-thick corona, plus a blackbody Comptonized by a hot and optically-thin corona. This model is similar to the understanding of the hard-state data of Aql X-1 by \citet{Sakurai12} and \citet{Sakurai14}, except that an inner part of the accretion disk is likely to be coverd by a cool corona. This double-Comptonization condition is possibly due to the rather high luminosity of this source.
\\
\\
The authors thank T. Hori for giving them the local model {\tt dkbbfth} and advice about it. This work was supported by the Grant-in-Aid for JSPS Fellows No. 15J08913.

\end{document}